# Empirical Evaluation of the Proposed eXSCRUM Model: Results of a Case Study

M. Rizwan Jameel Qureshi

**Faculty of Computing and Information Technology,**
**King Abdul Aziz University,**
**Ministry of Higher Education,**
**Jeddah, Kingdom of Saudi Arabia P.O.BOX 80221 Jeddah 21589**

**Abstract**
Agile models promote fast development. XP and Scrum are the most widely used agile models. This paper investigates the phases of XP and Scrum models in order to identify their potentials and drawbacks. XP model has certain drawbacks, such as not suitable for maintenance projects and poor performance for medium and large-scale development projects. Scrum model has certain limitations, such as lacked in engineering practices. Since, both XP and Scrum models contain good features and strengths but still there are improvement possibilities in these models. Majority of the software development companies are reluctant to switch from traditional methodologies to agile methodologies for development of industrial projects. A fine integration, of software management of the Scrum model and engineering practices of XP model, is very much required to accumulate the strengths and remove the limitations of both models. This is achieved by proposing an eXScrum model. The proposed model is validated by conducting a controlled case study. The results of case study show that the proposed integrated eXScrum model enriches the potentials of both XP and Scrum models and eliminates their drawbacks.

## 1. Introduction

Several researchers have discussed traditional development models during decades [1]. It is difficult to have a single common definition for traditional methodology. A traditional methodology is an explicit way of structuring one's thinking and actions [1]. Traditional methodologies are considered as heavyweight methodologies that are adopted for software development. In fact, traditional methodologies rely on a sequential series of steps that include requirements gathering, designing and building the solution, testing and deployment. In order to define and document, the traditional development methodologies, there is needed to establish consistent requirements from the start of a project. There are several development methodologies that include Waterfall, Spiral Model and Unified Process. In conventional software development methodologies, planning is done during the early stages of development that is strictly followed throughout development cycle. Finalizing the requirements during early phases may risk the success of a project. If there is no interaction of client with the development team during the development of a release/s, vague requirements may be considered by the team. This may leads to the failure of a project or may maximize the development cost. The traditional development methodologies stress on extensive documentation increasing the burden on development teams.

Due to above mentioned reasons, it is concluded that traditional software development methodologies cannot cope with the changing environment [2]. Therefore, new development methodologies were required to tackle the dynamically changing environment and requirements efficiently. Agile development methodologies include best software engineering practices that allows fast delivery of high quality software. The development approach is aligned with customer requirements and company objectives. The requirements and their solutions are built by collaboration of independent functional teams. Agile software development methodologies emphasize on direct interaction with the development team. For the distributed environment, the main modes of communications involve video conferencing, voice and e-mail.

Agile framework is based on iterative software development [3]. An independent working module is built after the completion of iteration. According to the authors [3], iteration must not consume more than two weeks to complete a code. Code is tested by a quality assurance team. The agile methodologies are light weight in nature [4]. Agile methodologies are suitable in changing environments because of new practices and principles helping to develop a product in short duration. XP model is one the most widely accepted agile models. Though agile XP model have several benefits but many software development companies hesitate to transit from traditional methodologies to agile XP model [5]. Main strengths of XP are fast development, cost saving, high satisfaction of client, test driven development resulting in less errors and acceptance of changing requirements. Following are few main limitations of XP model. XP model:

- focuses on code centered approach rather than design centered;
- recommends less documentation making it suitable only for small projects and limiting the opportunities and advantages of reusability;
- suggests to documenting the project after coding and this practice is very difficult and time consuming;
- lacks in structured reviews that ultimately results in lack of quality;
- Test driven approach is more time and cost consuming as compared to structured reviews.
- teams fully depend upon customer that may sometime become a cause for the failure of projects.

Scrum model is getting popularity from the last few years. Main strength of the Scrum model is high project management capability and its main limitations are summarized as follows.

- Scrum is a combination of generic project management practices and lacked in system development life cycle (SDLC) phases about engineering of a software
- As compared to XP model, Scrum demands high quality professionals to build scrum team
- Scrum lacks in the team activities to complete iterations in contrast to XP that has pair programming, continuous integration and automated builds.
- Lack of unit testing in SCRUM could lead to project failure.

Further paper is arranged as follows.
Section 2 focuses on the related work. Section 3 deals with the motivation towards the eXScrum. Section 4 proposes a new eXScrum model. Section 5 presents validation of the proposed new eXScrum model using a case study.

## 2. Related Work

A lot of research work has been done on agile development to explore and customize its practices since 2001. The main objectives of research are the issues raised during the implementation of phases and relevant agile principles and methods to be practiced. The authors [6] describe "knowledge engineering", a new approach of requirement analysis. This paper [7] focuses on the importance of understanding the requirements rather than jumping on the design and modeling. The importance of agility in requirement gathering and analysis is discussed in [7]. This research [8] describes agile techniques of requirement processes and proposed the "re-consideration" of electronic documentation of requirements in a project. A comparative analysis of requirement engineering between traditional software development methodologies and agile practices is described in [9]. The authors [9] also focused on advantages provides by the agile practices in requirement engineering that positively impact on the project velocity. The empirical study describes the agile requirement practices along with the advantages as well as challenges [10]. This research [11] described the advantages of agile requirement engineering and explored that how an "intranet" project adopt agile practices to save time and cost.
The authors [12,13,14,15,16] discussed the issues regarding architecture of agile models and their customization. The weaknesses in documentation phase of agile models are discussed to provide solutions [17,18,19,20,21]. These papers [22,23,24,25,26] focused on the testing issues using agile practices. Many researchers [27,28, 29,30] provide comparative analysis between agile software models and their customization like:

- experience of XP practices wrapped up with Scrum model;
- experience of satisfying the requirement of CMM level 2 and ISO 9001 with the combination of XP and Scrum models;
- experience of research context in building the agile software development group.

Next section describes the motivation for the eXScrum model to be proposed.

## 3. Motivation Towards eXScrum Model

A comparison of XP and SCRUM is provided in Table 1. The comparison is based on the quality parameters of agile practices and their level of availability in both models.

Table 1: A Comparison of XP and Scrum

| *Quality Parameter* | *XP* | *Scrum* |
|---|---|---|
| Engineering practices | Yes | No |
| Project management practices | No | Yes |
| Accept changes in iteration at any time | Yes | No |
| Requirement | Yes | No |

152

| | | |
|---|---|---|
| prioritization | | |
| Refactoring | Yes | No |
| Pair programming | Yes | No |
| Project size | Small to medium | Medium to high |
| Test driven development | Yes | No |
| Self organization | No | Yes |
| Unit testing | Yes | No |
| Design approach | Code centered | Design centered |
| Documentation level | Less | more |
| Team size | <10 | <10 and multiple teams |
| Code style | Clean and simple | Not specified |
| Technology Environment | Quick feed back | Not specified |
| Physical Environment | Co-located and limited distribution | Not specified |
| Business culture | Collaborative and cooperative | Not specified |
| Project size | Small to medium | Medium to high |
| Business culture | Collaborative and cooperative | Not specified |

Table 1 shows the main limitations of both XP and Scrum models. There is a desperate need of fusing the both XP and Scrum models to remove their shortcomings to solve the industry problem in major.

## 4. The Proposed eXScrum Model

The proposed eXScrum model is an improved paradigm enriching with complete project management and engineering practices of both Scrum and XP models. The main characteristic of the proposed eXScrum process model is that it provides a complete product development cycle without affecting the Scrum framework. All the engineering practices of the XP model exist in Sprint cycle of the Scrum model. Each phase of eXScrum model is shown in figure 1. Sprint zero is used before the start of the scrum development process. It is basically pre Scrum activity but in ordinary Scrum process, no clear guidelines or steps are defined. The Scrum model starts from product backlog. The proposed eXScrum process provides complete steps of the sprint zero. The process starts with the creation of product attributes and end resultant is product backlog. The proposed eXScrum model starts with product attributes which is much similar with user stories by the customer. Product attributes include salient features of a new product as required by the product master. Each item of the product attribute covers certain objectives as per client needs. Product attributes includes definitions of customer requirements that allows development team to produce a reasonable estimate of the effort during the implementation. Product attributes become part of the product backlog after going through the processes of estimation and prioritization.

The selected product attributes are estimated on the basis of effort required to build and implementing these attributes. The product attributes should be focused on user needs and benefits as opposed to specifying GUI layouts. The design focuses the current requirements. The formats of the design followed keep it simple (KIS) principle. In the designing phase, two types of diagrams are developed. These diagrams include class diagrams and object diagram. The class diagram is used to develop interfaces (front end), whereas object diagram helps in building database (back end). Test classes are also designed. The real development of the product is done during this phase. The process of coding requires coding standards, code ownership, pair programming and continuous integration.

Coding process needs continuous testing and refactoring. Code is tested frequently through unit tests. Each feature/attribute of the product is designed, implemented and tested individually with in the sprint development cycle. As a new code passes through testing, it is integrated to system. This process continues until the whole system is built. The process of continuous integration helps in reducing implementation risks. Continuous integration ensures that working module is available to use with new features. Scrum Meeting is conducted before start of the work daily. The duration of this meeting is 15 minutes. The main participants of this meeting are scrum master, product owner and development team.

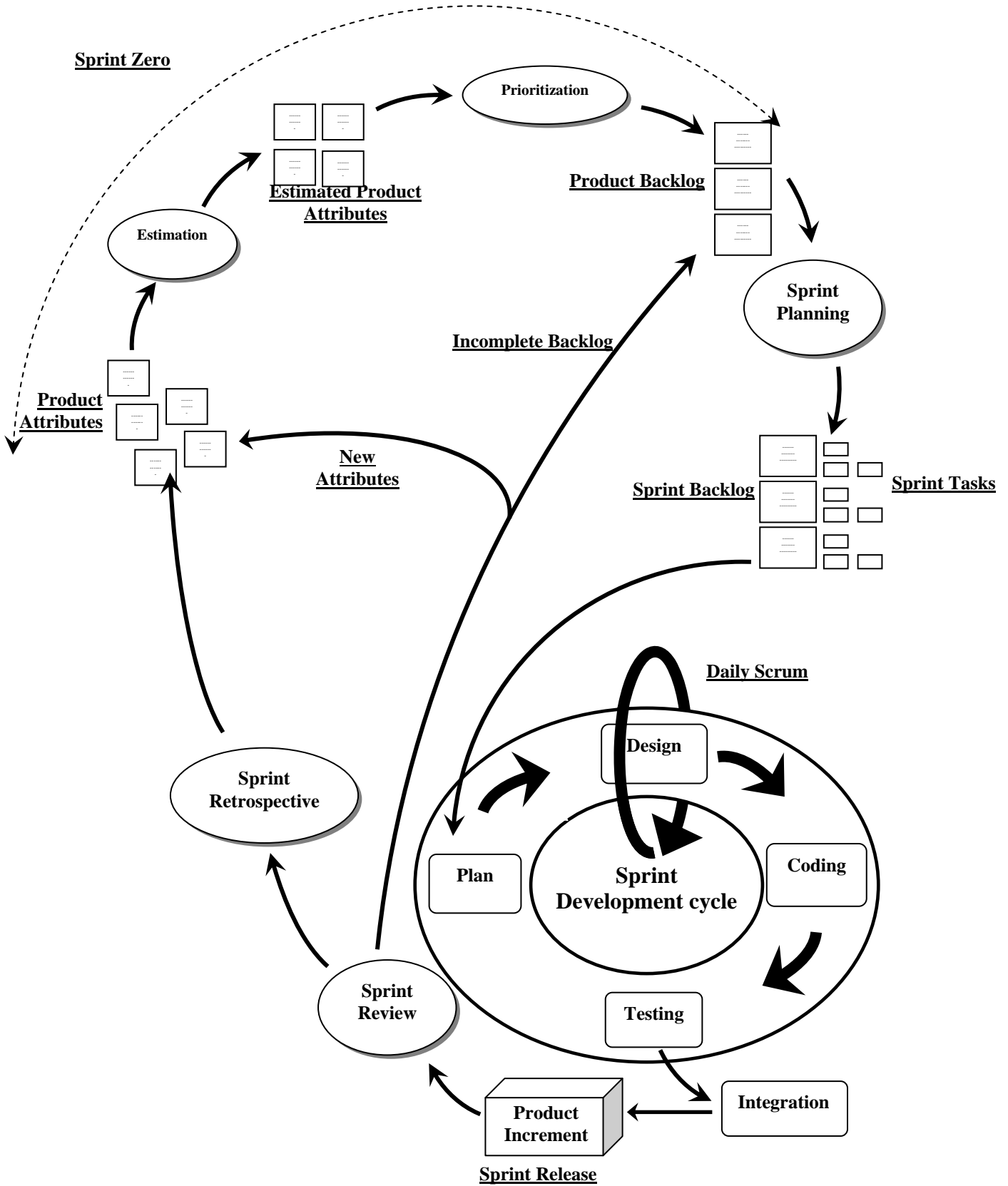

Fig. 1 The Proposed eXScrum Model

After the completion of sprint iteration, the working set of the product is released. This part of the product is presented in sprint review meeting. All stake holders are invited in the sprint review meeting. After the successful completion of all product increments, the whole product is launched with its full features. At the end of each sprint, sprint review meeting is conducted in which results of the new deliverable are provided to stakeholders. The purpose of Sprint review meeting is to ensure whether the required goals are achieved or not. The approval of the product increment depends upon the extent of the customer satisfaction.

The product owner is participated throughout the sprint development cycle of eXScrum model to achieve high level of customer satisfaction. The decision about next sprint backlog prioritization is taken in this meeting. The working of the product increment release is closely watched by all the stakeholders during the sprint review meeting. Any change or new suggestion, rises during sprint review meeting, becomes the part of sprint retrospective. The items of sprint retrospective become the part of next sprint. Sprint retrospective helps a team to be more successful to complete the next sprint. Only those items are considered in backlogs that do not disturb the normal working of the product increment. Although this process limits the working of product increment but it allows the opportunity to deliver a successful product increment at the end of each sprint.

## 5. Validation of the Proposed Model

The proposed eXScrum model is validated by conducting a case study to develop a payroll application for COMSATS Institute of information Technology Lahore, Pakistan. Manger account initiates the request for the development of Payroll Management System. A team was selected comprising of 6 members. The duration of project was five weeks. The description of case study project is provided in Table 2. The project was to complete in four iterations. An intensive one week training program was conducted to educate the team about the proposed model, agile XP and Scrum practices and principles before starting the project practically.

Table 2: Description of Case Study

| Characteristics | Description |
|---|---|
| Product Type | Payroll Application |
| Size | Medium |
| Project Type | Average |
| Type of Case Study | Controlled |

| Project Duration | 5 weeks |
|---|---|
| Iterations | 4 |
| Team size | 5 members |
| Programming Approach | Object Oriented |
| Feed back | Daily Feedback Require |
| Language | Java |
| Development Environment | Net Beans 6.9 |
| Documents | Ms Office XP |
| Other Tools | Rational Rose |
| Testing | J-Unit |
| Reports | IReport |
| Web Server | Apache Tomcat |

The main Scrum practices introduced to the team are sprint zero, product backlog, sprint backlog, sprint planning meetings, daily scrum meeting, sprint review meeting, and sprint retrospective. The main XP practices introduced to team during training are simple design, collective, pair programming, following coding standards, continuous testing, continuous integration and refactoring. Main tools used during the case study project are Rational Rose, Net Beans, My SQL, J-Unit, and IReport.

5.1 Empirical Analysis of the Case Study

The data is collected from four sprint releases. The collected data is represented in Table 3. All the columns represent cumulative/average data about releases of the case study while all the rows represent data of a particular attribute of the case study.

The first release (row one in the Table 3) was completed in two weeks time, whereas each of remaining three releases took one week duration. The term 'sprint release' is used in eXScrum model that shows the fact that system was released to actual customer test.

The number of modules (row 2), built during development process, are represented in each sprint release. Total tasks defined in these modules are represented in row 3. Each release shows number of tasks defined in their respective columns. Total work effort (row 4) of the project is remained constant throughout all releases. However, the direct hours dedicated to tasks (row 5) was reduced to 353 (h) in the first the release and in $2^{nd}$ and $3^{rd}$ release it was 150 (h) to 170 (h) respectively, whereas task effort was reduced to 120 (h) in $4^{th}$ release. Comparing task effort in percentage form (row 6), it was reduced from the initial 88% in 1st release to 75-85% in 2nd and 3rd releases respectively and was 60 % in 4th release. This indicates an increase in over-head for short development cycles.

Total number of interfaces built during the development was 48. The number of interfaces for the respective releases is represented in row 7. The line of code (LOC) of the interfaces in all sprint releases is 16820. Total number of classes built during the development process was 71 and line of code (LOC) of these classes remained 4240 (row 9 & 10). Total number of 24 test classes built for testing purpose having 8335 line of code (LOC).

The amount of logical lines of code, the team produced in a release, is represented in rows 11 and 12. Team's productivity (row 13) varied somewhat from 25.05 to 66.36 LOC/hour.

Test coverage is calculated as number of test LOC per system total LOC. Row 14 represents test coverage in percentage form. The results show that test coverage varies at each sprint release. The average test coverage as shown in row 13 is 51.81% of total LOC that is quite satisfactory. Row 15 shows integration data from the project which is used for software configuration management (SCM). The main purpose of software configuration management (SCM) is for tracking and controlling changes in the software.

Rows 16-17 are related to the quality of the system. The results show that total defect density of the system was relatively low. Post release defects per KLOC were 0.431. The defect density was evaluated as quite satisfactory that gives an indication of quality product.

Table 3: Exploratory Data from all Sprint Releases

| ID | Item | Release 1 | Release 2 | Release 3 | Release 4 | Total |
|---|---|---|---|---|---|---|
| 1 | Calendar Time (weeks) | 2 | 1 | 1 | 1 | 5 |
| 2 | Number of Modules (Items Sprint backlog) | 8 | 4 | 5 | 3 | 20 |
| 3 | Total Tasks defined | 50 | 12 | 14 | 6 | 82 |
| 4 | Total work effort (h) | 400 | 200 | 200 | 200 | 1000 |
| 5 | Task allocated actual hours | 353 | 150 | 170 | 120 | 793 |
| 6 | Task allocated actual (%) | 88 | 75 | 85 | 60 | 77 |
| 7 | Interfaces | 30 | 5 | 7 | 6 | 48 |
| 8 | Classes | 54 | 7 | 7 | 3 | 71 |
| 9 | Test Classes | 14 | 4 | 5 | 7 | 30 |
| 10 | Test Classes LOC | 11518 | 1780 | 2296 | 4182 | 19776 |
| 11 | Total LOC | 21036 | 3758 | 4365 | 7963 | 37122 |
| 12 | Total KLOC | 21.036 | 3.758 | 4.365 | 7.963 | 37.122 |
| 13 | Team Productivity (LOC/h) | 59.59 | 25.05 | 25.68 | 66.36 | 46.81 |
| 14 | Test Coverage (%) | 54.75 | 47.37 | 52.60 | 52.52 | 51.81 |
| 15 | Number of Integration | 40 | 22 | 30 | 20 | 112 |
| 16 | Post release defects | 7 | 3 | 3 | 3 | 16 |
| 17 | Post release defects /KLOC | 0.333 | 0.798 | 0.687 | 0.377 | 0.431 |

| 18 | Post release suggestions (Sprint Retrospective) | 7 | 5 | 4 | 1 | 17 |
| --- | --- | --- | --- | --- | --- | --- |
| 19 | Pair programming % | 80% | 80% | 80% | 80% | 80% |
| 20 | Customer involvement (sprint hrs/5 days (week)) | 30% | 28% | 20% | 22% | 25% |
| 21 | Customer Satisfaction | 80% | 80% | 90% | 90% | 85% |

In addition, 17 improvement suggestions are raised (row 18), i.e. new or improved user functionality. Most of the suggestions are raised from the first two releases.

Pair programming (row 19) was extensively exercised in the product development. The scheme of practicing pair programming was uniform throughout the product development i.e. 80 %.

In this controlled case study, the client shared the same office with the development team and thus was present over 80% of the total time, the actual customer involvement (row 20) was only 25% on average. This is a significant result since onsite customer is one of the most controversial topics in extreme programming methodology.

The customer satisfaction is measured in terms of satisfaction over number of modules of the product. The row 21 represents customer satisfaction in percentage form. Level of customer satisfaction remained at 80 % from sprint releases $1^{st}$ and $2^{nd}$ respectively. For releases $3^{rd}$ and $4^{th}$ the satisfaction level raised to 90 %.

## 6. Conclusion

Scrum model is one among the choices of agile methodologies that helps in managing projects efficiently. Scrum does not provide much more about how to engineer a product. XP model is also widely accepted model but there are few limitations in this model that need to be addressed. In this research, a novel eXScrum model is proposed that is an extended version of Scrum and XP models. The proposed model is validated by conducting a controlled case study. The results of case study show that the resultant integration enriches the potentials of both Scrum and XP models by eliminating their drawbacks.